\documentclass[showpacs,preprintnumbers,amsmath,amssymb,twocolumn]{revtex4}

\usepackage[normalem]{ulem}
\usepackage{graphicx}
\usepackage{dcolumn}
\usepackage{bm}
\usepackage{color}
\newcommand{\n}{\nonumber}

\newcommand{\la}{\langle}
\newcommand{\ra}{\rangle}
\newcommand{\rad}{\ra_{\cal D}}

\begin{document}

\title{Kullback--Leibler entropy and Penrose conjecture in the Lema\^{\i}tre--Tolman--Bondi model}

\author{Nan Li}
\email{linan@mail.neu.edu.cn}
\affiliation{College of Sciences, Northeastern University, Shenyang 110819, China}
\author{Xiao-Long Li}
\affiliation{College of Sciences, Northeastern University, Shenyang 110819, China}
\author{Shu-Peng Song}
\affiliation{College of Sciences, Northeastern University, Shenyang 110819, China}

\begin{abstract}
Our universe hosts various large-scale structures from voids to galaxy clusters, so it would be interesting to find some simple and reasonable measure to describe the inhomogeneities in the universe. We explore two different methods for this purpose: the Kullback--Leibler entropy and the Weyl curvature tensor. These two quantities characterize the deviation of the actual distribution of matter from the unperturbed background. We calculate these two measures in the spherically symmetric Lema\^{\i}tre--Tolman--Bondi model in the dust universe. Both exact and perturbative calculations are presented, and we observe that these two measures are in proportion up to second order.
\end{abstract}


\maketitle

\section{Introduction}

The standard model of cosmology is usually based on two preconditions: (1) the dynamics of cosmological evolution is governed by Einstein's general relativity; (2) the universe is spatially homogeneous and isotropic, as described by the Friedmann--Robertson--Walker (FRW) metric. The first precondition has stood many astronomical tests through the past century. However, the second one, always named ``cosmological principle'', is not so well established. Data from various cosmological experiments, have already confirmed with high precision that the universe is indeed very homogeneous at early times and large scales. However, at late times (matter-dominated era) or small scales ($10^2$ Mpc), due to gravitational instabilities, regions that are slightly overdense will attract matter from the surroundings, in the process becoming even more overdense and vice versa. As a result, the present universe has a well-developed nonlinear structure that cannot at all scales be described by the FRW model. Consequently, the cosmological principle still deserves serious considerations.

The irreversibility of structure formation in the universe reminds us of the process of entropy increasing in thermodynamics. Their resemblance naturally leads us to attempt to introduce some kind of ``entropy'' to characterize the cosmological structure formation. This issue aroused attention of many people in recent years. Motivated by the Penrose conjecture and thermodynamical considerations on gravitational field, Clifton, et al. \cite{Ellis} proposed a measure of gravitational entropy based on the Bel--Robinson tensor, which is the unique totally symmetric traceless tensor constructed from the Weyl tensor. It was shown that this measure is applicable to many models under certain conditions, ranging from the exact Schwarzschild black hole solution to the perturbed FRW model. Moreover, Sussman introduced a quasi-local scalar weighted average for the study of the Lema\^{\i}tre--Tolman--Bondi (LTB) dust model \cite{Sussman,Sussman2}. Considering the asymptotic limits in this framework, Sussman and Larena \cite{Larena} pointed out that the proposal in Ref. \cite{Ellis} is directly related to a negative correlation of the fluctuations of the energy density and the Hubble rate. Furthermore, these authors also explored the relative information entropy defined by Hosoya et al. \cite{entropy} and a variant based on the averaging procedure in Ref. \cite{Sussman}. They found that the independent proposals in Refs. \cite{Ellis,entropy} yield fairly similar conditions for entropy production, so they were able to obtain a robust qualitative inference of the evolution of gravitational entropies in Refs. \cite{Ellis,entropy} for the full evolution range of the generic LTB models.

The aim of this paper is to explore some quantity that can measure the structure formation in the inhomogeneous universe, i.e. to investigate some way to describe the deviation of the actual distribution of matter from the FRW background. This problem has been discussed in Ref. \cite{Li1}, where two different measures were studied: the Kullback--Leibler (KL) entropy ${\cal S}_{\cal D}$ and the full contraction of the Weyl tensor $C_{\mu\nu\lambda\rho}C^{\mu\nu\lambda\rho}$. It was found that they both serve as the reasonable measures for structure formation, and their only difference is a kinematical backreaction term ${\cal Q}_{\cal D}$,
\begin{align}
\frac{{{{\cal S}}}_{\cal D}}{V_{\cal D}}&=\frac{9}{32\pi}\left(\frac{t^2}{8}\la C_{\mu\nu\lambda\rho}C^{\mu\nu\lambda\rho}\rad+{\cal Q}_{\cal D}\right). \label{li}
\end{align}
We should emphasize that this relation was obtained in the linear cosmological perturbation theory \cite{Li1}. However, the three terms ${\cal S}_{\cal D}$, $C_{\mu\nu\lambda\rho}C^{\mu\nu\lambda\rho}$, and ${\cal Q}_{\cal D}$ are all of second order in the perturbative approach, so Eq. (\ref{li}) is their relation at the leading order. More calculational details will be discussed in Sect. \ref{sec:cal}.

The present paper is a successive research of Ref. \cite{Li1}. Here, we work in the spherically symmetric LTB model. We will prove that the kinematical backreaction vanishes in the perturbative approach in this model, with both growing and decaying modes of the scalar perturbations taken into account. Therefore, the KL entropy is actually proportional to the Weyl scalar in the LTB model,
\begin{align}
\frac{{{{\cal S}}}_{\cal D}}{V_{\cal D}}&=\frac{9t^2}{256\pi}\la C_{\mu\nu\lambda\rho}C^{\mu\nu\lambda\rho}\rad. \n
\end{align}
For simplicity, the gravitational constant $G$ is set to be unity throughout the paper.

\section{KL entropy, Weyl curvature, and kinematical backreaction}\label{sec:scq}

In this section, we explain the meanings of the KL relative entropy in information theory, the Weyl tensor in differential geometry, and the kinematical backreaction in the averaging problem in the inhomogeneous universe, respectively.

\subsection{KL Relative information entropy}\label{entr}

The relative information entropy in cosmology is a direct analog of the KL divergence widely used in statistics, probability theory, and information theory \cite{KL},
\begin{align}
S\{p||q\}:=\sum_{i}p_i\ln\frac{p_i}{q_i}, \n
\end{align}
which measures the difference between two probability distributions $\{p_i\}$ and $\{q_i\}$. Typically, $\{p_i\}$ denotes an actual distribution of data, while $\{q_i\}$ represents the presumed one or the theoretical description of $\{p_i\}$.

The KL divergence possesses several advantageous properties: (1) it is always nonnegative, $S\{p||q\}\geqslant 0$, with $S\{p||q\}$ $=0$ iff $p_i=q_i$; (2) it is invariant under parameter transformations; (3) it is additive for independent distributions; (4) it remains well defined for continuous distributions.

These properties inspired people to apply this idea to cosmology. In Ref. \cite{entropy}, Hosoya et al. defined the KL entropy density ${\cal S}_{\cal D}/V_{\cal D}$ as a functional of the actual and averaged distributions of mass densities, $\rho$ and $\la\rho\rad$, in the inhomogeneous universe,
\begin{align}
\frac{{\cal S}_{\cal D}\{\rho||\la\rho\rad\}}{V_{\cal D}}:=\left\la\rho\ln\frac{\rho}{\la\rho\rad}\right\rad, \label{KL}
\end{align}
where $V_{\cal D}$ is the volume of a domain $\cal D$. Furthermore, it proves that the increasing of ${\cal S}_{\cal D}$ encodes the non-commutation of temporal evolution and spatial averaging of the mass density \cite{entropy},
\begin{align}
\frac{{\dot{{\cal S}}}_{\cal D}}{V_{\cal D}}=\la\dot{\rho}\rad-\la\rho\rad^{^{\textbf{.}}}. \label{entropydot}
\end{align}

\subsection{Weyl curvature and Penrose conjecture} \label{weylt}

As matter and the geometry of space-time are closely interrelated in general relativity, it is also possible to depict the inhomogeneous distribution of matter via geometrical concepts. The idea was suggested by Penrose that the Weyl curvature tensor could play the role of gravitational entropy.

In differential geometry, the Weyl tensor $C_{\mu\nu\lambda\rho}$ is a measure of the curvature of a pseudo-Riemannian manifold, and in four-dimensional space-time, it is defined as
\begin{align}
C_{\mu\nu\lambda\rho}:=R_{\mu\nu\lambda\rho}-g_{\mu[\lambda}R_{\rho]\nu}+g_{\nu[\lambda}R_{\rho]\mu}+\frac13g_{\mu[\lambda}g_{\rho]\nu}R, \n
\end{align}
where $R_{\mu\nu\lambda\rho}$ is the Riemann tensor, $R_{\mu\nu}$ is the Ricci tensor, and $R$ is the Ricci scalar. We may regard the Weyl tensor as a part of the Riemann tensor, containing the components not captured by the Ricci tensor. Thus, it is locally independent of the energy-momentum tensor, so the Weyl tensor may be viewed as a purely geometrical description of an inhomogeneous space-time. Besides the same symmetries as the Riemann tensor, the Weyl tensor is traceless, ${C^\lambda}_{\mu\lambda\nu}=0$. Therefore, the full contraction of the Weyl tensor $C_{\mu\nu\lambda\rho}C^{\mu\nu\lambda\rho}$ is the principal scalar that we can construct.

For a Schwarzschild black hole of mass $M$, its Weyl scalar is $C_{\mu\nu\lambda\rho}C^{\mu\nu\lambda\rho}=12(2M)^2/r^6$. Meanwhile, the entropy $S$ of this black hole is $S=4\pi(2M)^2/4$. These observations led Penrose to conjecture that there could be some latent relationship between the thermodynamical entropy $S$ and the geometrical Weyl scalar $C_{\mu\nu\lambda\rho}C^{\mu\nu\lambda\rho}$ \cite{Penrose}. The various developments and modifications of Penrose's original conjecture can be found in Ref. \cite{wa}.

We may further ponder upon Penrose's conjecture in the evolution of the universe. In the early universe, when space-time is almost homogeneous, its Weyl tensor vanishes. But at late times, in the inhomogeneous space-time, the Weyl tensor will appear. Consequently, the averaged Weyl scalar $\la C_{\mu\nu\lambda\rho}C^{\mu\nu\lambda\rho}\rad$ plays the role of a measure for structure formation or a kind of entropy. For a more detailed discussion of Penrose's conjecture in cosmology, see Ref. \cite{Li1}.

\subsection{Averaging procedure}

We see in Sects. \ref{entr} and \ref{weylt} that ${\cal S}_{\cal D}$ and $\la C_{\mu\nu\lambda\rho}C^{\mu\nu \lambda\rho}\rad$ are both averaged quantities in the inhomogeneous universe. How to average a physical observable in the perturbed space-time is a long-standing and very complicated issue \cite{Ellisa}. However, for the objects with redshifts $\ll 1$, spatial averaging on a constant-time hyper-surface, for which the rest frames are complete, is already a good enough approximation.

In the following, we adopt the averaging formalism proposed by Buchert in Ref. \cite{bu}, and focus only on the scalars in the dust universe during the matter-dominated era. The metric of the inhomogeneous universe is written in the synchronous gauge as $ds^2=-dt^2+g_{ij}(t,{\bf x})\,dx^idx^j$, and volume average of a scalar $O(t,\bf x)$ in a comoving domain $\cal D$ on a constant-time hyper-surface is defined as
\begin{align}
\langle O \rad:=\frac{1}{V_{\cal D}(t)}\int_{\cal D} O(t,{\bf x})\sqrt{\det g_{ij}}\,d^3x, \n
\end{align}
where $V_{\cal D}(t):=\int_{\cal D}\sqrt{\det g_{ij}}\,d^3x$ is the volume of $\cal D$, and we may thus introduce an effective scale factor $a_{\cal D}(t)/a_{\cal D}(t_0):=(V_{\cal D}(t)/V_{\cal D}(t_0))^{1/3}$. For the perturbative calculations in Sect. \ref{sec:cal}, we further define the volume average on the spatially flat three-dimensional background as
\begin{align}
\la O\ra:=\frac{1}{\int_{\cal D} \,d^3x}\int_{\cal D} O(t,{\bf x})\,d^3x. \n
\end{align}

Applying Buchert's averaging procedure on the Einstein equations, we arrive at the generalized Friedmann equations for the irrotational dust universe \cite{bu},
\begin{align}
\left(\frac{\dot{a}_{\cal D}}{a_{\cal D}}\right)^2&=\frac{8\pi}{3}\la\rho\rad-\frac{\la {\cal R}\rad+{\cal Q}_{\cal D}}{6}, \n\\
\frac{\ddot a_{\cal D}}{a_{\cal D}}&=-\frac{4\pi}{3}\la\rho\rad+\frac{{\cal Q}_{\cal D}}{3}. \n
\end{align}
From these effective equations, we see that besides the ordinary entries in the Friedmann equations for the FRW model, two extra terms influence the evolution of the perturbed universe: the averaged three-dimensional spatial curvature $\la{\cal R}\rad$ and the so-called ``kinematical backreaction'',
\begin{align}
{\cal Q}_{\cal D}:=\frac{2}{3}(\la\theta^2\rad-\langle\theta\rad^2)-2\langle\sigma^2\rad. \label{Q}
\end{align}
${\cal Q}_{\cal D}$ bears this name because (1) it consists of kinematical quantities: the volume expansion scalar $\theta:={u^{\mu}}_{;\mu}$ and the shear scalar squared $\sigma^2:=\frac12\sigma_{\mu\nu}\sigma^{\mu\nu}$; (2) from Eq. (\ref{Q}), if ${\cal Q}_{\cal D}>0$, it plays the role of effective dark energy, and thus backreacts the evolution of the background universe.

\section{LTB model}\label{sec:LTB}

In this section, we first introduce the frequently used LTB model, and then calculate the three terms in Eq. (\ref{li}): ${\cal S}_{\cal D}$, $\la C_{\mu\nu\lambda\rho}C^{\mu\nu\lambda\rho}\rad$, and ${\cal Q}_{\cal D}$ in this model, respectively.

\subsection{LTB model and its solutions}\label{LTBm}

The LTB metric is an exact spherically symmetric (isotropic but maybe inhomogeneous) solution to Einstein's equations \cite{Krasinski}, which reads
\begin{align}
ds^2=-dt^2+\frac{R'(t,r)^2}{1+f(r)}\,dr^2+R(t,r)^2\,d\Omega^2, \label{LTB}
\end{align}
where $R(t,r)$ is a function of the cosmic time $t$ and the comoving radius $r$, and $f(r)>-1$ is an arbitrary function of $r$, with $f(r)/2$ being the energy per unit mass of the dust at the comoving radius $r$. In the following, we denote the partial derivative with respect to $t$ by $\dot{R}(t,r)$ and that to $r$ by $R'(t,r)$. It is obvious that if we further demand spatial homogeneity in this model, $R(t,r)=a(t)r$ and $f(r)=-kr^2$, the LTB metric reduces to the FRW model naturally.

Substitution of the LTB metric into Einstein's equations yields the dynamical equations for the dust universe,
\begin{align}
\frac{F'}{R^2R'}=8\pi\rho, \quad f=\dot{R}^2+2R\ddot{R}, \label{LTB1}
\end{align}
where
\begin{align}
F(r)=-2R^2\ddot{R}=\dot{R}^2R-fR \label{F}
\end{align}
is the second arbitrary function of $r$, with $F(r)/2$ denoting the mass within the sphere at the comoving radius $r$.

The solutions to $R(t,r)$ can be categorized into three classes: \\
(1) for $f=0$, the parabolic evolution,
\begin{align}
R=\left(\frac{9F}{4}\right)^{1/3}(t-T)^{2/3}, \label{1}
\end{align}
(2) for $f>0$, the hyperbolic evolution,
\begin{align}
R=\frac{F}{2f}(\cosh\eta-1), \quad t-T=\frac{F}{2f^{3/2}}(\sinh\eta-\eta), \label{2}
\end{align}
(3) for $f<0$, the elliptic evolution,
\begin{align}
R=\frac{F}{-2f}(1-\cos\eta), \quad t-T=\frac{F}{2(-f)^{3/2}}(\eta-\sin\eta), \label{3}
\end{align}
where $T=T(r)$ is the third arbitrary function of $r$, describing the time of big bang at the comoving radius $r$.

Furthermore, in any of the three cases above, the volume expansion scalar and the shear scalar squared are given as
\begin{align}
\theta=\frac{2{\dot R}}{R}+ \frac{\dot{R}'}{R'}, \quad \sigma^{2}=\frac{1}{3}\left(\frac{\dot{R}}{R}-\frac{\dot{R}'}{R'} \right)^{2}. \label{sigma}
\end{align}
These results will be used in the following calculations.

\subsection{Exact calculations in the LTB model}

Now, we review the exact results of ${\cal S}_{\cal D}$, $\la C_{\mu\nu\lambda\rho}C^{\mu\nu\lambda\rho}\rad$, and ${\cal Q}_{\cal D}$ in the LTB model. We calculate the time derivative of ${\cal S}_{\cal D}$ with the help of Eq. (\ref{entropydot}), instead of ${\cal S}_{\cal D}$, for mathematical convenience.

First, using Eqs. (\ref{entropydot}) and (\ref{LTB1}), we have the production rate of the KL entropy in the LTB model,
\begin{align}
\frac{\dot{\cal S}_{\cal D}}{V_{\cal D}}=&\frac{1}{8\pi}\left[\left\la\left(\frac{F'}{R^2R'}\right)^{\textbf{.}}\right\rad -\left\la\frac{F'}{R^2R'}\right\rad^{\textbf{.}}\right] \n\\
=&\frac{1}{4\pi G}\left[\left\la\left(\frac{2\dot{R}}{R} +\frac{\dot{R}'}{R'}\right)\left(\frac{2\ddot{R}}{R}+\frac{\ddot{R}'}{R'}\right)\right\rad \right. \n\\
&\left. -\left\la\frac{2\dot{R}}{R}+\frac{\dot{R}'}{R'}\right\rad\left\la\frac{2\ddot{R}}{R}+\frac{\ddot{R}'}{R'}\right\rad\right]. \label{sdot}
\end{align}

Second, the calculation of the Weyl scalar is straightforward. From the LTB metric in Eq. (\ref{LTB}), using Eq. (\ref{LTB1}), we have
\begin{align}
C_{\mu\nu\lambda\rho}C^{\mu\nu\lambda\rho}=\frac{16}{3}\left(\frac{\ddot{R}}{R}-\frac{\ddot{R}'}{R'}\right)^2. \label{Weylcon}
\end{align}
This result may be reexpressed in terms of the conformal Newman--Penrose scalar $\Psi_2$, which is related to the quasi-local density fluctuation (see Appendix D of Ref. \cite{Sussman}),
\begin{align}
\Psi_2=\frac{4\pi}{3}\rho-\frac{F}{2R^3}=\frac{1}{3}\left(\frac{\ddot{R}}{R}-\frac{\ddot{R}'}{R'}\right), \n
\end{align}
so the averaged Weyl scalar reads in a compact way,
\begin{align}
\la C_{\mu\nu\lambda\rho}C^{\mu\nu\lambda\rho}\rad=48\la[\Psi_2]^2\rad. \n
\end{align}

Last, from Eqs. (\ref{Q}) and (\ref{sigma}), we have the kinematical backreaction,
\begin{align}
{\cal Q}_{\cal D}=&\frac{2}{3}\left[\left\la\left(\frac{2\dot{R}}{R}+\frac{\dot{R}'}{R'}\right)^2\right\rad-\left\la\frac{2\dot{R}}{R}+ \frac{\dot{R}'}{R'}\right\rad^2\right. \n\\
&-\left.\left\la\left(\frac{\dot{R}}{R}-\frac{\dot{R}'}{R'}\right)^2\right\rad\right] \n\\
=&\left\langle  \frac{2\dot{R}}{R}\left(\frac{\dot{R}}{R}+\frac{2\dot{R}'}{R'} \right)\right\rangle_{\mathcal{D}}
-\frac23 \left\langle\frac{2\dot{R}}{R}+ \frac{\dot{R}'}{R'}\right\rangle_{\mathcal{D}}^{2}. \label{qdd}
\end{align}

The results in Eqs. (\ref{sdot})--(\ref{qdd}) are the exact expressions for ${\dot{\cal S}_{\cal D}}/{V_{\cal D}}$, $\la C_{\mu\nu\lambda\rho}C^{\mu\nu\lambda\rho}\rad$, and ${\cal Q}_{\cal D}$ in the LTB model. In general, it is highly nontrivial to work out some concise relation between them. This issue was extensively discussed in Refs. \cite{Sussman,Sussman2,Li1}.

In Ref. \cite{Morita}, numerical calculations were performed in a toy model to illustrate the evolutionary behaviour of ${\cal S}_{\cal D}$. Nevertheless, in order to gain quantitative relations, we still have to appeal to the perturbative approach. This will be the task in the next section.

\section{Perturbative calculations in the LTB model} \label{sec:cal}

In this section, we regard the LTB model as a spatially flat FRW model plus linear (first order) spherical perturbations. In this way, the three arbitrary functions $f(r)$, $F(r)$, and $T(r)$ in the LTB metric are solved as \cite{Morita1}
\begin{align}
f(r)&=\frac{20}{9}\psi'(r)r, \label{perf}\\
F(r)&=\frac{4}{9}r^3\left(1+\frac{10}{3}\psi(r)\right), \label{per}\\
T(r)&=-\frac{3}{2}\frac{\phi'(r)}{r}, \label{perT}
\end{align}
where $\psi(r)$ and $\phi(r)$ are the linear spherical scalar perturbations. A direct gauge transformation shows that $\psi=-\frac{9}{10}(\Psi+\frac16\Delta\chi)$ and $\phi=\frac{t}{2}[\chi+\frac{9}{5}t^{2/3}(\Psi+\frac16\Delta\chi)]$, where $\Psi$ and $\chi$ are the linear scalar perturbations in cartesian coordinate system in Ref. \cite{Li1}. Here, we should address that the linear regime around a spatially flat FRW model is necessarily restricted to early times and the decaying mode must be very subdominant (see numerical examples in Ref. \cite{Sussman2}). Besides, $\psi$ and $\phi$ can be mapped to the free parameters in the Hellaby--Lake conditions, which avoid the shell crossing singularities \cite{Hellaby}, and one of these conditions implies that $\phi'\leq 0$. This is consistent with the fact that the time of big bang $T(r)\geq 0$ at any comoving radius $r$. Last, by using the variables in Ref. \cite{Morita1}, it is difficult to identify an initial time slice that admits linearized initial conditions. However, other metric parametrization of the LTB metric, e.g. the one used in Refs. \cite{Sussman,Sussman2,Larena} may be more useful.

Below, we calculate ${\cal S}_{\cal D}/V_{\cal D}$, $\la C_{\mu\nu\lambda\rho}C^{\mu\nu\lambda\rho}\rad$, and ${\cal Q}_{\cal D}$ in the perturbative approach up to second order, but in fact only need to consult the first order perturbative results. This trick lies on the fact that all these three quantities are already of second order. We pick ${\cal S}_{\cal D}/V_{\cal D}$ for an example. If we expand $\rho$ to second order, $\rho=\rho^{(0)}+\rho^{(1)}+\rho^{(2)}$, we have
\begin{align}
\frac{{\cal S}_{\cal D}}{V_{\cal D}}=\left\la{\rho\ln\frac{\rho}{\la\rho\rad}}\right\rad=\frac{\la(\rho^{(1)})^2\ra-\la\rho^{(1)}\ra^2}{2\rho^{(0)}}+\cdots. \label{pS}
\end{align}
We see that the leading term in Eq. (\ref{pS}) is the variance of the mass density, and is thus of second order. Therefore, we are entitled to neglect the perturbation in $\sqrt{\det{g_{ij}}}$ and to use the average on the spatially flat three-dimensional background $\la\cdots\ra$ to replace the average in the perturbed space-time $\la\cdots\rad$, as their difference is at even higher orders. Similarly, this argument holds for $\la C_{\mu\nu\lambda\rho}C^{\mu\nu\lambda\rho}\rad$ and ${\cal Q}_{\cal D}$.

We should state here that since the three quantities ${\cal S}_{\cal D}$, $\langle C_{\mu\nu\lambda\rho}C^{\mu\nu\lambda\rho}\rangle_{\cal D}$, and ${\cal Q}_{\cal D}$ are all of second order and have no zeroth and first order terms, they are automatically gauge invariant quantities \cite{Stewart}, albeit the following calculations are performed in the synchronous gauge. An alternative gauge invariant treatment for the linear regime of the LTB models was furnished by the exact quasi-local perturbations in Ref. \cite{Sussman}, which has the advantage that it can track the perturbations through the nonlinear regime.

For the three solutions for $R$, we start from the simplest $f=0$ case, where there is only the decaying mode of the scalar perturbations. Next, we proceed to the growing mode in the $f\neq 0$ cases, and finally to the general case with both the decaying and growing modes taken into account.

\subsection{Decaying mode} \label{dm}

In the $f=0$ case, from Eq. (\ref{perf}), $\psi'=0$ and $\psi$ is a constant. Using Eqs. (\ref{1}) and (\ref{per}), we expand $R$ to first order,
\begin{align}
R(t,r)=rt^{2/3}\left(1+\frac{10}{9}\psi+\frac{\phi'}{rt}\right). \label{r1}
\end{align}
We see from Eq. (\ref{r1}) that the first two terms $1+10\psi/9$ are constant in time, and the third one $\phi'/(rt)$ represents a decaying mode in $R$. But this term should not be simply disregarded at present, because the constant perturbation $10\psi/9$ can be viewed as a fraction of the background metric and thus does not contribute to the perturbative results. This will be seen in Eqs. (\ref{s0})--(\ref{q0}).

Before giving the final results, two useful intermediate steps are listed below,
\begin{align}
\dot{R}&=\frac{2r}{3t^{1/3}}\left(1+\frac{10}{9}\psi-\frac{\phi'}{2rt}\right), \label{Rd}\\
R'&=t^{2/3}\left(1+\frac{10}{9}\psi+\frac{\phi''}{t}\right). \label{R'}
\end{align}
Substituting Eqs. (\ref{per}), (\ref{r1}), and (\ref{R'}) into Eq. (\ref{LTB1}), we obtain
\begin{align}
\rho=\frac{1}{6\pi t^2}\left(1-\frac{2\phi'}{rt}-\frac{\phi''}{t}\right). \n
\end{align}
Thus, we have the mass density at the background and first order,
\begin{align}
\rho^{(0)}(t)=\frac{1}{6\pi t^2}, \quad \rho^{(1)}(t,r)=-\frac{1}{6\pi t^3}\left(\frac{2\phi'}{r}+\phi''\right). \n
\end{align}
Substituting these results into Eq. (\ref{pS}), we attain the KL entropy in the LTB model up to second order,
\begin{align}
\frac{{\cal S}_{\cal D}}{V_{\cal D}}=\frac{1}{12\pi t^4}\left[\left\la\left(\frac{2\phi'}{r}+\phi''\right)^2\right\ra
-\left\la\frac{2\phi'}{r}+\phi''\right\ra^2\right]. \label{s0}
\end{align}
Above, we changed $\la\cdots\rad$ to $\la\cdots\ra$, as we have already explained.

Immediately, the time derivative and convexity of the KL entropy read
\begin{align}
\frac{\dot{\cal S}_{\cal D}}{V_{\cal D}}&=-\frac{1}{6\pi t^5}\left[\left\la\left(\frac{2\phi'}{r}+\phi''\right)^2\right\ra
-\left\la\frac{2\phi'}{r}+\phi''\right\ra^2\right], \label{sd0}\\
\frac{\ddot{\cal S}_{\cal D}}{V_{\cal D}}&=\frac{1}{2\pi t^6}\left[\left\la\left(\frac{2\phi'}{r}+\phi''\right)^2\right\ra
-\left\la\frac{2\phi'}{r}+\phi''\right\ra^2\right]. \label{sdd0}
\end{align}
From Eqs. (\ref{sd0}) and (\ref{sdd0}), we find that the KL entropy decreases (in a decelerated way) for the parabolic evolution in the LTB model. This result was exactly proven in Ref. \cite{sss}, and our perturbative calculation is consistent with this fact. Last, we should mention that if we start from Eq. (\ref{sdot}), we will arrive at the same result as that in Eq. (\ref{sd0}).

In like manner, substituting Eq. (\ref{r1}) into Eqs. (\ref{Weylcon}) and (\ref{qdd}), we get the averaged Weyl scalar,
\begin{align}
\la C_{\mu\nu\lambda\rho}C^{\mu\nu\lambda\rho}\rad=\frac{64}{27t^6}\left\la\left(\frac{\phi'}{r}-\phi''\right)^2\right\ra, \label{c0}
\end{align}
and the kinematical backreaction,
\begin{align}
{\cal Q}_{\cal D}=&\frac{2}{3t^4}\left[\left\la\left(\frac{2\phi'}{r}+\phi''\right)^2\right\ra-\left\la\frac{2\phi'}{r}+\phi''\right\ra^2\right. \n\\
&\left.-\left\la\left(\frac{\phi'}{r}-\phi''\right)^2\right\ra\right]. \label{q0}
\end{align}
Above, we did not combine the first and third terms in ${\cal Q}_{\cal D}$, as we notice that the third term exactly cancels the averaged Weyl scalar (up to a coefficient). We may discover from Eq. (\ref{q0}) that ${\cal Q}_{\cal D}$ just characterizes the difference between ${\cal S}_{\cal D}/V_{\cal D}$ and $\la C_{\mu\nu\lambda\rho}C^{\mu\nu\lambda\rho}\rad$.

From Eqs. (\ref{s0}), (\ref{c0}), and (\ref{q0}), we may formally express the relation of ${\cal S}_{\cal D}/V_{\cal D}$, $\la C_{\mu\nu\lambda\rho}C^{\mu\nu\lambda\rho}\rad$, and ${\cal Q}_{\cal D}$ as
\begin{align}
\frac{{\cal S}_{\cal D}}{V_{\cal D}}=\frac{9}{32\pi G}\left(\frac{t^2}{8}\la C_{\mu\nu\lambda\rho}C^{\mu\nu\lambda\rho}\rad+\frac{4}{9}{\cal Q}_{\cal D}\right). \n
\end{align}
This result looks rather like that in Eq. (\ref{li}), but with a different coefficient: $4/9$. However, there is no problem. In Ref. \cite{Li1}, merely the growing mode of the linear perturbations were considered, but here we see from Eq. (\ref{r1}) that only the decaying mode exists in the $f=0$ case, which was not extensively discussed in Ref. \cite{Li1}.

Nevertheless, in the $f=0$ case, ${\cal Q}_{\cal D}$ is actually found to vanish. This is because
\begin{align}
{\cal Q}_{\cal D}=\frac{2}{3t^4}\left[3\left\la\frac{\phi'}{r} \left(\frac{\phi'}{r}+2\phi''\right)\right\ra-\left\la\frac{2\phi'}{r}+\phi''\right\ra^2\right], \n
\end{align}
and the integrals above can be performed directly. Consider a spherical domain with the comoving radius $R_{\cal D}$, for a second order quantity $O$, we have
\begin{align}
\la O\ra=\frac{4\pi}{V_{\cal D}}\int_0^{R_{\cal D}}OR^2R'\,dr=\frac{3}{R_{\cal D}^3}\int_0^{R_{\cal D}}Or^2\,dr, \n
\end{align}
where $R^2R'=r^2t^2$ on the spatially flat three-dimensional background, and $V_{\cal D}=4\pi\int_0^{R_{\cal D}}r^2t^2\,dr=\frac{4\pi}{3}R_{\cal D}^3t^2$. In this way, it is easy to find
\begin{align}
3\left\la\frac{\phi'}{r} \left(\frac{\phi'}{r}+2\phi''\right)\right\ra=\left\la\frac{2\phi'}{r}+\phi''\right\ra^2=\frac{9\phi'(R_{\cal D})^2}{R_{\cal D}^2}. \n
\end{align}
This proves ${\cal Q}_{\cal D}=0$. In fact, in the parabolic evolution, ${\cal Q}_{\cal D}$ vanishes exactly. A direct proof can be found in Ref. \cite{ps}, and a general discussion was shown in Sect. 5.1 in Ref. \cite{sss}.

Eventually, we obtain
\begin{align}
\frac{{\cal S}_{\cal D}}{V_{\cal D}}=\frac{9t^2}{256\pi}\langle C_{\mu\nu\lambda\rho}C^{\mu\nu\lambda\rho}\rangle_{\cal D}. \label{s1}
\end{align}
We see that the relation of ${\cal S}_{\cal D}/V_{\cal D}$ and $\la C_{\mu\nu\lambda\rho}C^{\mu\nu\lambda\rho}\rad$ is more direct in the LTB model, as ${\cal Q}_{\cal D}$ vanishes in this case. Furthermore, if we multiply $V_{\cal D}$ on both sides of Eq. (\ref{s1}), we find that the total KL entropy in a domain ${\cal D}$ is the same as the total Weyl scalar. But we should stress that this integral equality does not mean that Eq. (\ref{s1}) holds pointwise, and Eq. (\ref{s1}) holds only in the perturbative approach.

\subsection{Growing mode} \label{gm}

For the cases with a non-vanishing $f$, we first Taylor expand $t-T$ in Eqs. (\ref{2}) and (\ref{3}), solve the parameter $\eta$, and then substitute it into the corresponding $R$. For both cases $f>0$ and $f<0$, we arrive at the same result,
\begin{align}
R(t,r)=rt^{2/3}\left(1+\frac{10}{9}\psi+\frac{\phi'}{rt}+\frac{\psi't^{2/3}}{r}\right). \label{r2}
\end{align}
This result is the same as that in Eq. (\ref{r1}), but with an additional term ${\psi't^{2/3}}/{r}$, because $\psi'$ is now nonzero if $f\neq 0$. This term is the growing mode in the perturbative expansion of $R$, and will dominate in $R$ as $t$ increases. For this reason, we may first neglect the decaying mode $\phi'/(rt)$ in Eq. (\ref{r2}), and focus on the growing and constant ones,
\begin{align}
R(t,r)=rt^{2/3}\left(1+\frac{10}{9}\psi+\frac{\psi't^{2/3}}{r}\right). \label{r2g}
\end{align}
Now, $R$ is a function of $\psi$ only.

The following perturbative calculations are totally parallel to those in Sect. \ref{dm}. First, we have
\begin{align}
\rho^{(1)}=-\frac{1}{6\pi t^{4/3}}\left(\frac{2\psi'}{r}+\psi''\right). \n
\end{align}
Then ${\cal S}_{\cal D}/V_{\cal D}$, $\dot{{\cal S}}_{\cal D}/V_{\cal D}$, and $\ddot{{\cal S}}_{\cal D}/V_{\cal D}$ are obtained in order,
\begin{align}
\frac{{\cal S}_{\cal D}}{V_{\cal D}}&=\frac{1}{12\pi t^{2/3}}\left[\left\la\left(\frac{2\psi'}{r}+\psi''\right)^2\right\ra-\left\la\frac{2\psi'}{r}+\psi''\right\ra^2\right],\label{sg} \\
\frac{\dot{{\cal S}}_{\cal D}}{V_{\cal D}}&=\frac{1}{9\pi t^{5/3}}\left[\left\la\left(\frac{2\psi'}{r}+\psi''\right)^2\right\ra-\left\la\frac{2\psi'}{r}+\psi''\right\ra^2\right],\label{sgg} \\
\frac{\ddot{{\cal S}}_{\cal D}}{V_{\cal D}}&=\frac{1}{27\pi t^{8/3}}\left[\left\la\left(\frac{2\psi'}{r}+\psi''\right)^2\right\ra-\left\la\frac{2\psi'}{r}+\psi''\right\ra^2\right].\label{sggg}
\end{align}
From Eqs. (\ref{sgg}) and (\ref{sggg}), we see that the KL entropy in the LTB model increases monotonically (in an accelerated way) both for the hyperbolic and elliptic evolutions. This agrees with the result in Ref. \cite{Li1}, though seems to disagree with that in Eq. (\ref{sd0}). But actually, there is no contradiction, as Ref. \cite{Li1} only dealt with the growing modes in the perturbed universe. For more non-perturbative analyses on the temporal evolutions of the KL entropy, see Ref. \cite{sss}.

Furthermore, we have $\la C_{\mu\nu\lambda\rho}C^{\mu\nu\lambda\rho}\rad$ and ${\cal Q}_{\cal D}$ as
\begin{align}
\la C_{\mu\nu\lambda\rho}C^{\mu\nu\lambda\rho}\rad=\frac{64}{27t^{8/3}}\left\la\left(\frac{\psi'}{r}-\psi''\right)^2\right\ra \label{cg}
\end{align}
and
\begin{align}
{\cal Q}_{\cal D}=&\frac{8}{27t^{2/3}}\left[\left\la\left(\frac{2\psi'}{r}+\psi''\right)^2\right\ra-\left\la\frac{2\psi'}{r}+\psi''\right\ra^2\right.\n\\
&\left. -\left\la\left(\frac{\psi'}{r}-\psi''\right)^2\right\ra\right]. \label{qg}
\end{align}

From Eqs. (\ref{sg}), (\ref{cg}), and (\ref{qg}), we formally recover the result in Eq. (\ref{li}),
\begin{align}
\frac{{{{\cal S}}}_{\cal D}}{V_{\cal D}}&=\frac{9}{32\pi}\left(\frac{t^2}{8}
\la C_{\mu\nu\lambda\rho}C^{\mu\nu\lambda\rho}\rad+{\cal Q}_{\cal D}\right). \label{hhh}
\end{align}
Till now, we understand that this relation is valid only for the growing mode of the scalar perturbations. As expected, the second order perturbative calculations in the LTB model for the $f\neq 0$ cases reconfirm this relation.

However, as we demonstrate in the $f=0$ case, ${\cal Q}_{\cal D}$ also vanishes in the $f\neq 0$ cases. In this way, we again have
\begin{align}
\frac{{\cal S}_{\cal D}}{V_{\cal D}}=\frac{9t^2}{256\pi}\langle C_{\mu\nu\lambda\rho}C^{\mu\nu\lambda\rho}\rangle_{\cal D}. \n
\end{align}
This result is the same as that in Eq. (\ref{s1}).

\subsection{General case} \label{general}

With the preparations in Sects. \ref{dm} and \ref{gm}, we now present the general solutions for ${{\cal S}_{\cal D}}/{V_{\cal D}}$, $\la C_{\mu\nu\lambda\rho}C^{\mu\nu\lambda\rho}\rad$, and ${\cal Q}_{\cal D}$, taking into account both the decaying and growing modes of the scalar perturbations. We begin with
\begin{align}
R(t,r)=rt^{2/3}\left(1+\frac{10}{9}\psi+\frac{\phi'}{rt}+\frac{\psi't^{2/3}}{r}\right), \n
\end{align}
and obtain the full expressions for the KL entropy,
\begin{align}
\frac{{\cal S}_{\cal D}}{V_{\cal D}}=&\frac{1}{12\pi t^4}\left[\left\la\left(\frac{2\phi'}{r}+\phi''+\frac{2\psi't^{5/3}}{r}+\psi''t^{5/3}\right)^2\right\ra\right. \n\\
&-\left.\left\la\frac{2\phi'}{r}+\phi''+\frac{2\psi't^{5/3}}{r}+\psi''t^{5/3}\right\ra^2\right], \n
\end{align}
the averaged Weyl scalar,
\begin{align}
&\la C_{\mu\nu\lambda\rho}C^{\mu\nu\lambda\rho}\rad \n\\
=&\frac{64}{27t^6}\left\la\left(\frac{\phi'}{r}-\phi''+\frac{\psi' t^{5/3}}{r}-\psi''t^{5/3}\right)^2\right\ra, \n
\end{align}
and the kinematical backreaction,
\begin{align}
{\cal Q}_{\cal D}=&\frac{8}{27t^4}\left[\left\la\left(\frac{3\phi'}{r}+\frac32\phi''-\frac{2\psi't^{5/3}}{r}-\psi''t^{5/3}\right)^2\right\ra\right. \n\\ &-\left\la\frac{3\phi'}{r}+\frac{3}{2}\phi''-\frac{2\psi't^{5/3}}{r}-\psi''t^{5/3}\right\ra^2 \n\\
&-\left.\left\la\left(\frac{3\phi'}{2r}-\frac{3}{2}\phi''-\frac{\psi't^{5/3}}{r}+\psi''t^{5/3}\right)^2\right\ra\right]. \n
\end{align}

These are the final and complete results for ${\cal S}_{\cal D}/V_{\cal D}$, $\la C_{\mu\nu\lambda\rho}C^{\mu\nu\lambda\rho}\rad$, and ${\cal Q}_{\cal D}$ that we hope to calculate in this paper, with all the perturbative modes considered. After some algebra, the seemingly complicated ${\cal Q}_{\cal D}$ proves to vanish in the perturbative approach again, and we eventually arrive at the relation between the KL entropy and the Weyl scalar in the general case,
\begin{align}
\frac{{\cal S}_{\cal D}}{V_{\cal D}}=\frac{9t^2}{256\pi}\langle C_{\mu\nu\lambda\rho}C^{\mu\nu\lambda\rho}\rangle_{\cal D}. \n
\end{align}

\section{Conclusions and discussions} \label{sec:con}

In recent years, the study of the inhomogeneous cosmological models and the corresponding problems, e.g. the averaging procedure, backreaction mechanism, and light propagation in perturbed space-time, has attracted much attention (see Refs. \cite{attention1,attention2} and the references therein). One relevant and important issue is to seek some simple and reasonable measure for the large-scale structure formation during cosmological evolution. In Ref. \cite{Li1}, two such measures were investigated: the KL entropy ${\cal S}_{\cal D}$ and the averaged Weyl scalar $\la C_{\mu\nu\lambda\rho}C^{\mu\nu\lambda\rho}\rad$, and their relation is shown in Eq. (\ref{li}) in the perturbative approach up to second order. In the present paper, we verify this result in the LTB model, and simultaneously point out that the kinematical backreaction vanishes in this special model, due to its higher symmetry. Consequently, there is a more concise relation between the KL entropy and the averaged Weyl scalar---they are in proportion in the LTB model,
\begin{align}
\frac{{\cal S}_{\cal D}}{V_{\cal D}}=\frac{9t^2}{256\pi}\langle C_{\mu\nu\lambda\rho}C^{\mu\nu\lambda\rho}\rangle_{\cal D}. \n
\end{align}
This result applies to all the three types of evolution in the LTB model (up to second order).

Finally, we give some general discussions.

(1) The exact results for the KL entropy, Weyl scalar, and kinematical backreaction in the LTB model are listed in Eqs. (\ref{sdot})--(\ref{qdd}). However, their fully nonlinear exact relationship is still under consideration and this seems to be a highly nontrivial task. A next possible step should be to look for other quantities that vanish in the perturbative treatment but are present in the full LTB solution. In this aspect, the non-perturbative quasi-local averaging formalism in Refs. \cite{Sussman,Larena} provided a reasonable approach, which differs from Buchert's procedure but coincides in the linear regime around a spatially flat FRW background. In this formalism, the authors were able to express the KL entropy, Weyl scalar, kinematical backreaction, and other tensorial objects in terms of the quadratic fluctuations of the density and the Hubble expansion scalar (see Sects. 6 and 7 in Ref. \cite{Sussman}).

(2) From Eqs. (\ref{sigma}) and (\ref{Weylcon}), we find $C_{\mu\nu\lambda\rho}C^{\mu\nu\lambda\rho}\propto\sigma^2$. This is not just a coincidence, and we may have deeper insight from this proportion. The Weyl curvature may be irreducibly decomposed into the electric part $E_{\mu\nu}:=C_{\mu\lambda\nu\rho}u^\lambda u^\rho$ and the magnetic part $H_{\mu\nu}:=\frac12\epsilon_{\mu\lambda\alpha\beta}{C^{\alpha\beta}}_{\nu\rho}u^\lambda u^\rho$. In the LTB model (both in exact and perturbative approaches), the magnetic part vanishes, so $C_{\mu\nu\lambda\rho}C^{\mu\nu\lambda\rho}\propto E_{\mu\nu}E^{\mu\nu}$. At the same time, the shear tensor is proportional to the electric part, $\sigma^2\propto E_{\mu\nu}E^{\mu\nu}$. These facts explain the similar results in Eqs. (\ref{sigma}) and (\ref{Weylcon}).

(3) To our knowledge, the Penrose conjecture has not yet been well formulated in a rigorous mathematical way. Hence, to construct possible scalars from the Weyl tensor should be the first step in this direction. According to the Petrov classification, in addition to $C_{\mu\nu\lambda\rho}C^{\mu\nu\lambda\rho}$, there are other independent full contractions, e.g. $\epsilon_{\mu\nu\lambda\rho}C^{\lambda\rho\alpha\beta}{C_{\alpha\beta}}^{\mu\nu}$, $C_{\mu\nu\lambda\rho}C^{\lambda\rho\alpha\beta}{C_{\alpha\beta}}^{\mu\nu}$, or $\epsilon_{\mu\nu\lambda\rho}C^{\lambda\rho\alpha\beta}C_{\alpha\beta\gamma\delta}C^{\gamma\delta\mu\nu}$. A direct calculation indicates that $\epsilon_{\mu\nu\lambda\rho}C^{\lambda\rho\alpha\beta}{C_{\alpha\beta}}^{\mu\nu}$ vanishes. In Ref. \cite{Goode:1992pp}, it was shown that $C_{\mu\nu\lambda\rho}C^{\mu\nu\lambda\rho}$ diverges and thus fails to be monotonic near the isotropic singularities. Therefore, some other candidates have been considered, e.g. \linebreak $(C_{\mu\nu\lambda\rho}C^{\mu\nu\lambda\rho})/(R_{\mu\nu} R^{\mu\nu})$, which may help to evade this limitation, and this direction deserves further exploration.

(4) From a mathematical point of view, the curvature of space-time is measured by the Riemann tensor, consisting of the Ricci tensor and Weyl tensor, namely, ${\rm Riemann}={\rm Ricci}+{\rm Weyl}$. However, Einstein's equations only associate the Ricci sector with the energy-momentum tensor. We may naturally ask why the information stored in the Weyl sector is absent in general relativity. A possible answer is that the Weyl tensor is linked not to the dynamical, but to the thermodynamical aspect of gravitational fields. The evolution of our universe is doubtlessly irreversible. But on the contrary, a process governed by Einstein's equations possesses the invariance of time reversal, so the time asymmetry of cosmological evolution is not shown explicitly in Einstein's equations. Is this information encoded in the Weyl tensor? Penrose proposed that some scalar invariant of the Weyl tensor could be identified with the gravitational entropy of the universe. Our present work helps to confirm this idea and indicates that the Weyl tensor can be further related to the KL entropy. These facts lead us to wonder whether there exist equations that are parallel to Einstein's equations and quantify the thermodynamical relationship between space-time and matter. These equations are expected to link the Weyl tensor with the concepts such as temperature or entropy. This question will be the topic of research in future.

\vskip .3cm

We thank T. Buchert, M. Morita, and D. J. Schwarz for fruitful discussions. We are very grateful to the anonymous referee for the detailed and valuable comments. This work is supported by the National Natural Science Foundation of China (No. 11105026) and the Fundamental Research Funds for the Central Universities (No. N140504008).

\end{document}